\documentclass[mypaper,8pt,twoside]{CoAst}
\usepackage{epsf,graphicx,fancyhdr}
\renewcommand{\chaptermark}[1]{\markboth{#1}{}}

\newcommand{\kopf}{\small\itshape Comm. in Asteroseismology \\ Contribution to the Proceedings of the Wroclaw HELAS Workshop, 2008}

\newcommand{\Authors}[1]{\begin{center}\normalsize\bf\sf #1 \end{center}}

\renewcommand{\author}[1]{\begin{center}\normalsize\bf\sf #1 \end{center}}
\newcommand{\Address}[1]{\begin{center}\small\sf #1 \end{center}}

\newcommand{\Session}[1]{{\vspace{3mm}\small \noindent  \hspace*{3mm} Session: } #1 \normalsize}

\newcommand{\Objects}[1]{{\vspace{0mm}\small \noindent  \hspace*{3mm} Individual Objects: } \small #1 \normalsize}

	\newcommand{\poster}{\small Poster \newline}

\renewenvironment{abstract}{\section*{Abstract}\normalsize\sf}{}
\newcommand{\References}[1]{\begin{flushleft}{\large References\\}\vspace*{2mm}\small #1 \end{flushleft}}

\newcommand{\chapterCoAst}[2]{\chapter[\sf\normalsize #1\\ \footnotesize \hspace*{5mm}by #2 \sf\normalsize][]{#1\\}\rhead[\fancyplain{}{\sf\footnotesize \center{#1}}]{\fancyplain{}{\sffamily\thepage}}\lhead[\fancyplain{\kopf}{\sffamily\thepage}]{\fancyplain{\kopf}{\sf\footnotesize \center{#2}}}}

\renewcommand{\chaptername}{}

\newcommand{\acknowledgments}[1]{\vspace*{5mm}\noindent  \textbf{Acknowledgments.} #1}

\def\rfr{\smallskip\par\noindent
        \hangindent=7truemm
        \hangafter=1}

\begin{document}
\sf

\chapterCoAst{Pulsations, chemical composition and multiplicity in main-sequence A- and F-type stars}
{S.\,Hekker, Y.\,Fr\'emat, P.\,Lampens, P\, De Cat} 
\Authors{S.\,Hekker, Y.\,Fr\'emat, P.\,Lampens, P.\, De Cat} 
\Address{Royal Observatory of Belgium, Ringlaan 3, 1180 Brussels, Belgium}

\noindent
\begin{abstract}
The region in the HR-diagram where the main sequence intersects the classical instability strip hosts A- and F-type stars exhibiting a rich variety of physical phenomena, such as pulsations on various time scales and chemical peculiarities. We aim to investigate the occurrence of these phenomena among suspected binary systems in this region of the HR-diagram and their mutual interactions.
\end{abstract}

\Session{ \poster } 
\Objects{$\sim$ 65 A and F main sequence stars} 

\section*{Introduction}
Main-sequence A and F stars are interesting since stars in the same region of the HR-diagram show (different types of) pulsations and/or chemical peculiarities. The necessary (different) conditions for these phenomena to occur in stars of similar luminosity and temperature are not yet (well) known. For a long time it has been thought that chemical peculiarity and pulsations were mutually exclusive (e.g. Breger 1970). Radiative diffusion (gravitational settling of Helium and levitation of other elements in layers which are sufficiently stable against turbulent mixing) in slowly rotating stars can explain the existence of chemical peculiarities as well as the suppression of pulsations due to a lack of Helium in the partial ionisation zone which drives the pulsations. By now, stars possessing both chemical peculiarities and pulsations are observed, which can not be fully understood in terms of the radiative diffusion hypothesis. Only for some subgroups of chemically peculiar pulsating stars theoretical explanations exist. We refer to e.g. Kurtz (2000) for an extensive review. 

The aim of our project is to investigate empirically the limits where chemical peculiarity and pulsations can co-exist, where they are mutually exclusive and the role of multiplicity therein.

\section*{Sample selection and observations}
A sample of poorly studied A and F stars is selected to investigate the mutual interactions between pulsations, multiplicity and chemical abundances in the region where the classical instability strip intersects the main sequence. Stars are selected to be brighter than $8^{th}$ magnitude in V (Hipparcos catalogue: ESA 1997), have a variability flag in the radial velocity catalogue of Grenier et al. (1999) and only a few references in Simbad.

During three observing runs using the ELODIE spectrograph mounted on the 1.93-m telescope at the Observatoire de Haute-Provence, France, spectra with S/N $\geq$ 80 were collected for about 65 stars. To be able to search for pulsations and/or multiplicity, the observing strategy was such that both long- and short-term radial velocity variations could be identified, i.e., several exposures (exposure times $\leq$ 10 min) during a single night and at least one during another night were gathered. 

\section*{Determination of stellar parameters and chemical abundances}
Stellar parameters are determined using GIRFIT a computer code which performs least squares fitting based on the MINUIT minimisation package (see also Fr\'emat et al. 2007). The radial velocity (RV), projected rotational velocity ($v \sin i$), effective temperature (T$_{\rm{eff}}$) and surface gravity ($\log$~g) are derived in four consecutive steps: (1) RV is determined from the cross-correlation functions; (2) $v \sin i$ is based on metallic line fitting with RV fixed; (3) T$_{\rm{eff}}$ is derived from H$_{\alpha}$, with $\log$~g equal to 4.0; (4) $\log$~g is derived by computing the luminosity using the parallax, V magnitude and T$_{\rm{eff}}$ and, from these, the mass and radius from stellar evolutionary tracks (Schaller et al. 1992).
Because $\log$~g depends on T$_{\rm{eff}}$ the last two steps are performed iteratively.

Determination of abundances is performed for elements with strong spectral lines based on solar abundances in the range 4500 - 5500 \AA , using synthetic spectra computed with the SYNSPEC computer code (Hubeny \& Lanz 1995) and least squares fitting based on MINUIT. The strategy is as follows: (1) microturbulence is determined from several sensitive isolated lines; (2) iron abundance is determined using strong iron lines (EW $>$ 5 m\AA); (3) abundances of iron peak elements (Ti, V, Cr, Mn, Co, Ni, Cu) are determined from isolated lines or lines blended with iron; (4) using the previous results, iron peak elements are re-determined also allowing blends of several of these elements in the same fitting interval; (5) abundances of other elements with single lines are determined.
The abundance determination is performed semi-automatically and optimisation of the procedure is still ongoing.

\section*{Current status \& Future prospects}
So far, the cross-correlation functions are computed and checked for indications of pulsations and multiplicity. Indeed, in this rather poorly studied sample of stars binary systems, pulsators and stars without obvious radial velocity variations are present. For the single stars in our sample the determination of stellar parameters is nearly finished. At the moment the abundance determination as described in the previous section is being tested on simulated data as well as on reference stars also observed with the ELODIE spectrograph. Detailed information on the developed procedures and results will be published in subsequent publications.


\acknowledgments{SH acknowledges financial support from the Belgian Federal Science Policy (ref: MO/33/018). This research has made use of the SIMBAD database operated at CDS, Strasbourg, France.}

\References{
\rfr Breger M. 1970, ApJ, 162, 597
\rfr Fr\'emat Y., Lampens P., Van Cauteren P. et al. 2007, A\&A, 471, 675
\rfr Grenier S., Baylac M.-O., Rolland L. et al. 1999, A\&AS, 137, 451
\rfr Hubeny I. \& Lanz T. 1995, ApJ, 439, 2, 875
\rfr Kurtz D.W. 2000, ASPC, 210, 287
\rfr ESA 1997, ESA SP, 1200,1
\rfr Schaller G., Schaerer D., Meynet G. \& Maeder A. 1992, A\&AS, 96, 2, 269
}

\end{document}